\documentclass[12pt, prd, showpacs,superscriptaddress,notitlepage]{revtex4-1}

\usepackage[a4paper, margin=0.8in]{geometry}
\usepackage{setspace}

\usepackage{amssymb}
\usepackage{amsmath}

\usepackage{graphicx}

\usepackage{color}
\definecolor{darkblue}{rgb}{0,0,1}
\definecolor{darkgreen}{rgb}{0,.4,0}
\definecolor{deepred}{rgb}{.5,0,0}
\usepackage[backref,pagebackref=true, colorlinks,linkcolor=darkblue,urlcolor=deepred,citecolor=darkgreen,unicode]{hyperref}

\begin{document}

\title{Pressure of thermal excitations in superfluid helium}
\author{I. N. Adamenko}
\author{K.E. Nemchenko}
\affiliation{Karazin Kharkov National University,
         Svobody Sq. 4, Kharkov, 61077, Ukraine.}
\author{I. V. Tanatarov}
\email{igor.tanatarov@gmail.com}
\affiliation{Akhiezer Institute for Theoretical Physics,
        NSC KIPT of NASU, Academicheskaya St. 1, Kharkov, 61108, Ukraine.}
\author{A.F.G.~Wyatt}
\affiliation{School of Physics, University of Exeter,
          Exeter EX4 4QL, UK.}

\begin{abstract}

We find the pressure, due to the thermal excitations of superfluid helium, at
the interface with a solid. The separate contributions of phonons, $R^-$ rotons
and $R^+$ rotons are derived. The pressure due to $R^-$ rotons is shown to be
negative and partially compensates the positive contribution of $R^+$ rotons,
so the total roton pressure is positive but several times less than the
separate $R^-$ and $R^+$ roton contributions. The pressure of the quasiparticle
gas is shown to account for the fountain effect in $HeI\!I$.  An experiment is
proposed to observe the negative pressure due to $R^-$ rotons.
\end{abstract}
\maketitle

\section{Introduction}

The physical characteristics of superfluid $^4$He can often be well described
in terms of the elementary
excitations from a coherent ground state, which acts as a vacuum state for the
excitations. The elementary excitations are phonons at low temperatures and
both phonons and rotons at higher temperatures. In thermal equilibrium, the
excitations form  a gas of quasiparticles which behaves in a similar way to a
gas of atoms. When the excitations are incident on an interface with a solid,
they reflect and create a pressure. However there are differences from a gas of
atoms; the dispersion law is very different and there are excitations with a
negative group velocity. Also excitations in the liquid helium can create
phonons in the solid, and vice versa. The vapour pressure due to phonons and
rotons and quantum evaporation, was considered in  \cite{Wyatt1984}.

\begin{figure}[t]
\begin{center}
\includegraphics[viewport=90 260 500 580, width=0.6\textwidth]{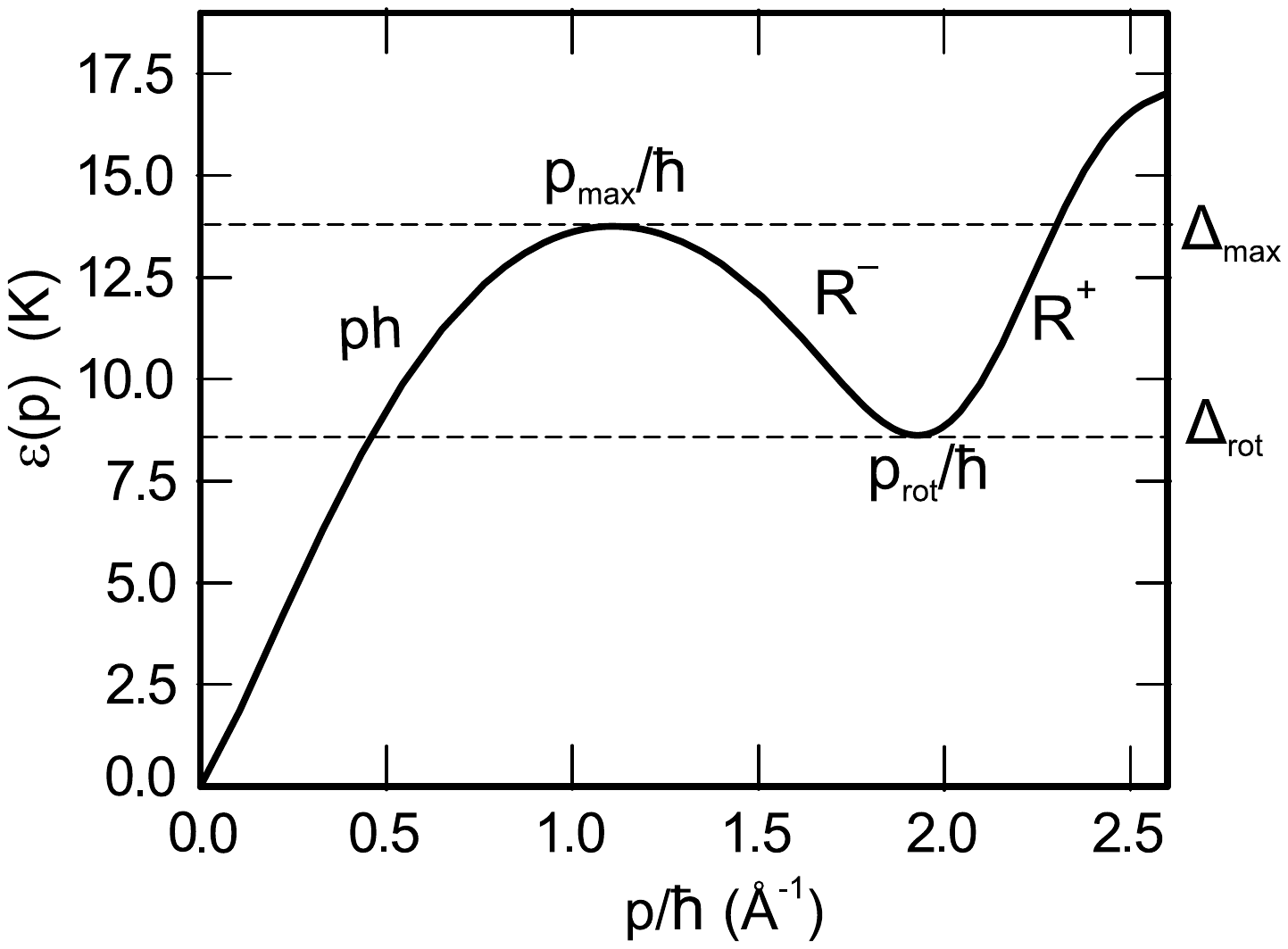}
\parbox{0.9\textwidth}
	{\caption{\label{Fig1}The measured dispersion curve of superfluid helium.
The phonon, $R^-$ roton and $R^+$ roton regions are shown.}}
\end{center}
\end{figure}

The dispersion curve of superfluid helium is non-monotonic and consists of
three monotonic parts. Because of this the gas of quasiparticles of superfluid
helium has three quite distinct components, phonons, $R^-$ rotons and
$R^+$ rotons, each component corresponding to a monotonic region of the curve,
see Fig.\ref{Fig1} and Ref.\cite{neutron}. The $R^-$ rotons are described by the
descending part of the dispersion curve and have negative group velocity, so
that their momentum is directed opposite to the direction of propagation. There
are many indirect experiments that verify the quasiparticle description of
thermal
fluctuations in He~II. Moreover, there are also experiments \cite{exp1,exp2},
in which phonons and rotons are directly detected. The $R^-$ rotons
were shown to have a
negative group velociy, relative to their momentum, in \cite{R-reg}.

In this paper we obtain the pressure of superfluid helium quasiparticles on the
boundary with another continuous medium. The processes that underly the
interaction of the quasiparticles with an interface, their reflection,
transmission and creation, are considered self-consistently in a
general approach. We take into account the possibility of the
conversion of quasiparticles into each other, on reflection from the
interface, and derive the
separate contributions of phonon, $R^-$ roton and $R^+$ roton modes to the full
pressure of quasiparticles on the interface.

We start with a short section  on the thermodynamics of the gas of superfluid
helium quasiparticles and derive an expression for the pressure. In
the next section we consider momentum
transfer to the interface due to different quasiparticles
interacting with it. We derive the total pressure of quasiparticles
on the interface, as
well as the individual contributions of phonon, $R^-$ roton and $R^+$
roton modes
to it. We show that no detailed expressions for the  probabilities of
the processes are
necessary; the principle of detailed balance is all that is needed to
derive the pressure.
The pressure can be expressed in the form for the pressure of a
classical gas on an
interface, although the meaning is different.
  The part of the pressure due to $R^-$ roton is shown to be
negative. It is compensated by the positive part of $R^+$ rotons, and the total
positive pressure is less than the absolute values of either the
$R^-$ or $R^+$
rotons.

These results should be applicable for any two adjacent continuous media
at low temperatures; in particular we see that the resultant force, with which
quasiparticles of both media are acting on the interface, is directed from the
medium with lesser sound velocity to the medium with the greater one. We also
show that the expression for the total pressure is the same as that
derived from
thermodynamics.

We then discuss contributions to the pressure from the $R^+$ rotons
and $R^-$ rotons
as calculated by thermodynamics and kinematics. The two theories give
very different results and it is clear that
the kinematic approach is correct.
Hence it is impossible to calculate their separate contrbutions to
the fountain pressure from their separate entropies.

In the last section we suggest an experiment that would measure the negative
pressure that the $R^-$ rotons exert on the boundary with a solid.

\section{Pressure of quasiparticles from thermodynamics}

The dispersion curve of superfluid helium $\varepsilon(p)$  has the
maxon maximum $\Delta_{max}$ at $p\!=\!p_{max}$ and roton minimum
$\Delta_{rot}$ at $p\!=\!p_{rot}$, see Fig.\ref{Fig1}.
So for the range of energies $E\!\in\!(\Delta_{rot},\Delta_{max})$ there are
three real roots of equation $\varepsilon(p)\!=\!E$, which corresponding to
phonons, $R^-$ rotons, and $R^+$ rotons. We number the roots in the ascending
order of their absolute values $p_{i}(\varepsilon)$:
$p_{3}\!>\!p_{rot}\!>\!p_{2}\!>p_{max}\!>\!p_{1}\!>\!0$. Thus $i\!=\!1$
corresponds to phonons, $i\!=\!2$ to $R^-$ rotons, and $i\!=\!3$ to $R^+$
rotons.

The thermodynamics of He~II quasiparticles starts from the Helmholtz
free energy, per unit volume, of superfluid helium (see for example
\cite{Khalatnikov}), which
can be written as a sum $F\!=\!\sum_{i}F_{i}$, where
\begin{equation}
     \label{F}
     F_{i}=-k_{B}T\int\!
\frac{d^{3}p}{(2\pi\hbar)^{3}}\ln\left[1+n_{T}(\varepsilon_{i}(p))\right]\quad
\mbox{for}\;i\!=\!1,2,3
  \end{equation}
  are the phonons and $R^{\pm}$ rotons contributions to $F$. Here $n_{T}$ is the
Bose-Einstein distribution and $k_{B}$ Boltzmann constant. The
integration domains
of the integrals are $p_{1}\!\in\!(0,p_{max})$, $p_{2}\!\in\!(p_{min},p_{max})$
and $p_{3}\!\in\!(p_{min},p_{\infty})$.

The analytical expressions are obtained with the help of the usual
approximations of the dispersion relation: linear for phonons
$\varepsilon(p_{1})\!=\!sp_{1}$ ($s$ is the sound velocity of liquid
helium), and
parabolic for rotons
$\varepsilon(p_{2,3})\!=\!k_{B}\Delta_{rot}+(p_{2,3}\!-\!p_{rot})^{2}/2\mu$
($\mu$ is the so-called roton mass). Usually the $R^-$ and $R^+$ roton's
contributions are calculated together, and one obtains (see for example
\cite{Khalatnikov} or \cite{Wilks})
\begin{eqnarray}
     \label{phononF}
     F_{ph}&=&
     -\frac{\pi^2}{90}\frac{(k_{B}T)^4}{\hbar^{3}s^{3}}\\
     \label{rotonF}
     F_{rot}&=&
     -\frac{p_{rot}^{2}\sqrt{\pi\mu/2}}{2\pi^{2}\hbar^{3}}
         (k_{B}T)^{3/2}
      e^{-\Delta_{rot}/T}.
\end{eqnarray}

The pressure $P$ of a system is given by
\begin{equation}
P=-\Big( \frac{\partial F}{\partial V}\Big)_T
\end{equation}
so as Eqs. (\ref{phononF}) and (\ref{rotonF}) are the free energies
per unit volume,
they are the negative of the pressure.
Hence the total pressure $P$ due to phonons and rotons is:
\begin{equation}
     \label{pressure}
     P=
    \frac{\pi^2}{90}\frac{(k_{B}T)^4}{\hbar^{3}s^{3}}\\
      +\frac{p_{rot}^{2}\sqrt{\pi\mu/2}}{2\pi^{2}\hbar^{3}}
         (k_{B}T)^{3/2}
      e^{-\Delta_{rot}/T}.
\end{equation}

As this pressure is temperature dependent, it causes  the fountain effect,
which was observed by Allen and Jones
\cite{Allen1938}.
Consider two reservoirs of liquid helium at temperatures $T$ and $T+\delta T$,
and pressures $P$ and $P+\delta P$, respectively.
Hence

\begin{equation}
\delta P=-\Big(\frac{\partial F}{\partial T}\Big)_V  \delta T=S \delta T
\end{equation}
where $S$ is the entropy per unit volume. The second term is the
usual expression for
the fountain pressure given by London \cite{London}.  It  can  also
be  obtained
     from    the    equation    for   the superfluid   velocity
$\mathbf{v}_{s}$,      in      the      two-fluid      model,     with
$\mathbf{v}_{s}\!=\!0$ \cite{Tilly}.

If we treat the phonons and rotons in a naive way, and consider them
as a classical gas (without any
justification),
the classical formula for the pressure of a gas of
for each component, gives
\begin{equation}
      \label{Pclass}
      P_{i}(T)=\frac{1}{3}\int \frac{d^{3}p_{i}}{(2\pi\hbar)^{3}}
      n_{T}\!(\varepsilon(p_{i}))p_{i}u_{i}(p_{i})\quad\mbox{for}\;i\!=\!1,2,3.
\end{equation}
Here $u_{i}\!=\!d\varepsilon(p_{i})/dp_{i}$ are group velocities of
quasiparticles. Due to the negative dispersion of $R^-$ rotons, their
contributions then should be negative. Although Eq. (\ref{Pclass}) is valid for
classical particles, it is not justified for phonons and rotons which
do not just
reflect from the surface but have other possibilities
  such as a mode change. For example a
phonon can change into $R^+$ roton. However, this classical  approach
suggests that the
negative dispersion of $R^-$ rotons will lead to a negative presure
contribution.

To calculate the negative pressure of the $R^-$ rotons we use a
kinematic approach in the next section.
We derive the
contributions of phonons, $R^-$ and $R^+$ rotons to pressure
by calculating the momentum transferred to the helium-solid interface
when quasiparticles of different types are incident on it.
We also calculate the pressure of a
quasiparticle beam incident on an interface which cannot be obtained
from thermodynamics.

\section{Kinematic calculation of the pressure of quasiparticles}
\subsection{Momentum transferred to the interface}

Let there be a flat boundary $z\!=\!0$ between superfluid helium and a solid.
The solid has a monotonic dispersion relation and occupies the region
$z\!<\!0$, and superfluid helium with its non-monotonic dispersion relation
$\varepsilon(p)$ fills the region $z\!>\!0$.  For simplicity we take into
account only the longitudinal phonons in the solid, to which we assign index
$i\!=\!4$.  Generalising the problem, to take into account the
presence of both longitudinal
and transverse phonons, is trivial, as can be seen below, and actually does not
affect the results.

\begin{figure}[t]
\begin{center}
\includegraphics[viewport=112 306 483 536, width=0.6\textwidth]{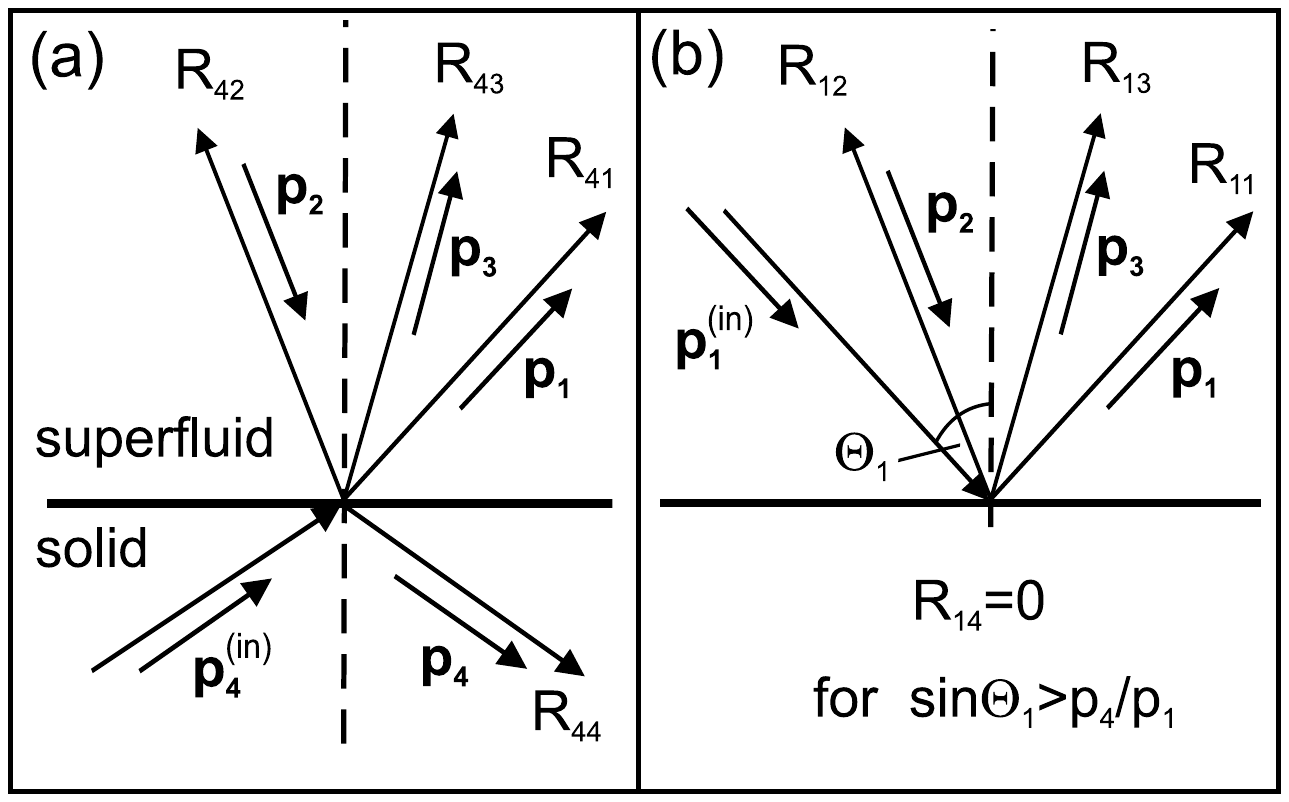}
\parbox{0.9\textwidth}
	{\caption{\label{Fig2}Two examples of an incident quasiparticle creating other quasiparticles at the interface between the superfluid and a solid. In (a) the incident quasiparticle is a phonon in the solid and in (b) the incident quasiparticle is a phonon in the superfluid. The $\textrm{R}_{ij}$ are the probabilities of creating the out-going quasiparticle. The arrows labelled $\textbf p_i$ show the direction of the momenta of the quasiparticles.}}
\end{center}
\end{figure}

When any quasiparticle, of energy
$\varepsilon\!\in\!(\Delta_{rot},\Delta_{max})$, is incident on the interface
it is destroyed, and a quasiparticle of one of the four types
$i\!=\!1,2,3,4$ is
created on the interface. There are three possibilities in the helium
($i\!=\!1,2,3$) and one in the
solid ($i\!=\!4$). The analogous situation appears in ordinary acoustics, when the incident phonon with linear dispersion may with some probability be reflected (i.e. it is destroyed and the reflected phonon is created), or with some probability may turn into the phonon of the other medium. Our case is different because the dispersion relation of He~II is nonlinear and nonmonotonic, so the quasiparticle created in helium can be of any of the three possible types.
Consider a quasiparticle of type $i$ with momentum
$\mathbf{p}_{i}$ and energy $\varepsilon(\mathbf{p}_{i})$  be incident on the
interface with the solid and a quasiparticle of type $j$ ($j\!=\!1,2,3,4$) is
created with probability  $R_{ij}$. Two examples are shown in Fig.\ref{Fig2}. 

Energy is conserved in these processes. Due to the translational
invariance of the
interface, the transverse component of momentum $\mathbf{p}_{\tau}$ is also
conserved. So $\varepsilon$ and $\mathbf{p}_{\tau}$ are set by the incident
quasiparticle; the momenta of the other quasiparticles are
$p_{i}(\varepsilon)$, and
the angles of propagation $\Theta_{i}$, measured from the normal, are found
from $p_{\tau}\!=\!p_{i}(\varepsilon)\sin\Theta_{i}$ for $i\!=\!1,2,3,4$. Here
$p_{4}\!=\!p_{sol}(\varepsilon)$ is determined by the dispersion relation of
the solid.

The normal component of momentum of the incident quasiparticle is
$p_{iz}^{(in)}$, and the created quasiparticles are $p_{jz}$ for
$j\!=\!1,2,3,4$. The created quasiparticles travel away from the interface, but
the momentum of the $R^-$ roton is towards the interface. Then the signs of
$p_{iz}$, for the case
$p_{iz}$ are real, are
\begin{equation}
\label{rot-}
p_{1z},p_{3z}>0,\quad p_{2z},p_{4z}<0.
\end{equation}

For channel $j$ a momentum $\Delta p_{ij}$ is transferred to the interface and
hence to liquid or solid as a whole. It is found from momentum conservation
\begin{equation}
     \label{momentum}
         \Delta p_{ij}= p_{iz}^{(in)}-p_{jz}.
  \end{equation}
  On summing $\Delta p_{ij}$ over the channels $j\!=\!1,2,3,4$ with their
probabilities $R_{ij}$, taking into account that for the incident wave
$p_{iz}^{(in)}\!=\!-p_{iz}$, and using the normalizing condition
  \begin{equation}
     \label{normalization}
     \sum\limits_{j=1}^{4}R_{ij}=1\quad \mbox{for}\; i=1,2,3,4,
  \end{equation}
we  obtain
\begin{equation}
      \label{momFull}
      \Delta p_{i}=\sum\limits_{j=1}^{4}\Delta p_{ij}R_{ij}=
      -p_{iz}-\sum\limits_{j=1}^{4}R_{ij}p_{jz}.
\end{equation}
Here $\Delta p_{i}$ is the average momentum transferred to the interface per
one incident  quasiparticle of type $i$.

Let the incident quasiparticles be in thermodynamic equilibrium, with the
Bose-Einstein distribution $n_{T}(\varepsilon)$. Then the momentum
transferred to the interface, per unit time per unit area, by the incident
quasiparticles
of type $i$ is
\begin{equation}
      \label{PI}
      \Pi_{i}\!=\!-
          \int\!\frac{d^{3}p_{i}}{(2\pi\hbar)^{3}}
          n_{T}\!(\varepsilon)|u_{i}|\cos{\Theta_{i}}
          \left\{
          p_{iz}\!+\! \sum\limits_{j=1}^{4}R_{ij}p_{jz}
          \right\}.
\end{equation}
Here $u_{i}$ is the group velocity of quasiparticle $i$, $u_{i}\!=\!u(p_{i})$
for $i\!=\!1,2,3$ and $u_{4}\!=\!u_{sol}$, so that
$n_{T}\!(\varepsilon)|u_{i}|\cos{\Theta_{i}}$ is the number of quasiparticles
$i$ incident on the interface per unit time per unit area. The domain of
integration is $\Theta_{i}\!\in\!(0,\pi/2)$, and the probabilities $R_{ij}$ are
nonzero only when both $p_{iz}$ and $p_{jz}$ are real, so the momentum
$\Pi_{i}$ is real.

The principle of detailed balance can be written in terms of probabilities as
functions of the conserved quantities, $\varepsilon$ and $p_{\tau}$:
\begin{equation}
     \label{DetBalance}
     R_{ij}(\varepsilon,p_{\tau})=R_{ji}(\varepsilon,p_{\tau})\quad\mbox{for}\;
i,j=1,2,3,4.
\end{equation}
  It is convenient to change the variables of integration in Eq. (\ref{PI}) to
the arguments of $R_{ij}(\varepsilon,p_{\tau})$. Then from Eq. (\ref{PI})
\begin{equation}
      \label{PIe-p}
      \Pi_{i}\!=\!-
      \int\frac{d\varepsilon}{8\pi^{2}\hbar^{3}}
      n_{T}\!(\varepsilon)\!\!
      \int\! dp_{\tau}^{2}
      \left\{
          p_{iz}\!+\!\sum\limits_{j=1}^{4}R_{ij}p_{jz}
      \right\}.
\end{equation}
The integral of the sum should be treated as the sum of integrals, with domains
of
integration in which the integrands are real.

The full momentum transferred to the interface per unit time per unit area by
all the incident quasiparticles, $\Pi$, is the sum of Eq. (\ref{PIe-p}) over
$i\!=\!1,2,3,4$. With the help of Eqs. (\ref{normalization}) and
(\ref{DetBalance}), we obtain
\begin{equation}
     \label{!}
     \sum\limits_{i=1}^{4}\left\{
     p_{iz}+\sum\limits_{j=1}^{4}R_{ij}p_{jz}
     \right\}
     =2\sum\limits_{i=1}^{4}p_{iz}.
\end{equation}
  So, on summing, we see that the total pressure does not depend on the
probabilities $R_{ij}$ and thus on the reflection properties of the interface.
The terms that remain, $p_{iz}$, reflect the bulk properties of the two
adjacent media, i.e. their dispersion relations.

Taking into account the signs of $p_{iz}$ (\ref{rot-}), we have
$p_{iz}\!=\!p_{i}\mbox{sgn}(p_{iz})\cos\Theta_{i}$, and then on taking the
interior integral, we obtain
\begin{equation}
      \label{P3D}
      \Pi(T)
      =\sum\limits_{i=1}^{4}\Pi_{i}=-
      \int\frac{d\varepsilon}{6\pi^{2}\hbar^{3}}
      n_{T}\!(\varepsilon)\sum\limits_{i=1}^{4}p_{i}^{3}(\varepsilon)\,\mbox{sgn
}
(p_{iz}).
\end{equation}

\subsection{Pressure of the quasiparticles  and the contribution of\\ the three modes.}

The quantity $\Pi$ in Eq. (\ref{P3D}) is the difference between the pressure of
the  phonons in the solid $P_{sol}\!=\!P_{4}$ where $z\!<\!0$ and the pressure
of the quasiparticles in the superfluid $P\!=\!P_{1}\!+\!P_{2}\!+\!P_{3}$ where
$z\!>\!0$: $\Pi\!=\!P_{sol}-P$. The contributions from phonons ($i\!=\!1$),
$R^-$
rotons ($i\!=\!2$), and $R^+$ rotons ($i\!=\!3$) to the total pressure of the
quasiparticles in the superfluid are
\begin{equation}
      \label{Pi3D}
      P_{i}(T)=\mbox{sgn}(p_{iz})
      \int\frac{d\varepsilon}{6\pi^{2}\hbar^{3}}
      n_{T}\!(\varepsilon)p_{i}^{3}(\varepsilon)\quad\mbox{for}\;i=1,2,3.
\end{equation}
The pressure of the longitudinal phonons of the solid $P_{4}$ is given by
relation (\ref{Pi3D}) for $i\!=\!4$, in which the signum is assumed value $+1$.
The transverse phonons would  give additional pressure
$P_{5}(T)$, given also by relation (\ref{Pi3D}) with the corresponding
dispersion relation and positive sign, but we shall see later that it
is unnecessary to include them.

Changing the integration variables back to $d^{3}p_{i}$, we obtain that the
expressions for each of the contributions of quasiparticles $i$ for
$i\!=\!1,2,3,4$ take the universal form of the gas-kinetic equation for the
pressure of a gas of classical particles (\ref{Pclass}).

We see that the contribution $P_2$ of the $R^-$ rotons is \textit{negative}.
The reason is their negative group velocity $u_{2}\!<\!0$, which determines the
sign of $p_{2z}$ in Eqs. (\ref{rot-}) and (\ref{Pi3D}) and enters explicitly in
Eq. (\ref{Pclass}). The contribution $P_2$ is composed of the momentum that the
incident $R^-$ rotons bring to the interface (which is negative), as well as
the negative
momentum transferred to the interface on creation of the $R^-$ rotons by either
phonons, or rotons, or phonons in the solid. The latter terms are proportional
to coefficients $R_{i2}$ and are summed up in such a way that the coefficients
disappear, and the expression for the full pressure (\ref{Pi3D}) takes the form
that could be obtained by considering (wrongly) that all the quasiparticles are
just specularly
reflected without mode change or transmission.

Hence we have proved the classical formula (\ref{Pclass}) to be applicable for
the gas of quasiparticles, for which mode changes on interaction with the
interface are allowed. However, it should be emphasized that one should
distinguish the pressure of quasiparticles of type $i$ incident on the
interface, which is given by $\Pi_{i}$ of Eq. (\ref{PI}) and essentially
depends on explicit functions $R_{ij}$, from the contribution of the mode $i$
to full pressure $P\!=\!P_{1}\!+\!P_{2}\!+\!P_{3}$, given by definition by $P_{i}$ of Eq. (\ref{Pi3D}) or (\ref{Pclass}), which does not depend on $R_{ij}$.

In order to obtain Eq. (\ref{Pi3D}) and (\ref{Pclass}), we did not make any
assumptions on the explicit forms of either dispersion relation $\varepsilon
(p)$ or the probabilities $R_{ij}$, so the result (\ref{Pclass}) should be
applicable to any two continuous media with a common boundary (see for example
\cite{Portnoi}). As a consequence, from Eq. (\ref{Pi3D}) we see that
quasiparticles of the medium with lesser sound velocity act on the interface
with greater pressure than those of the medium with greater sound velocity. If
there is more than one type of quasiparticles in either of the media, the
corresponding contributions to pressure from the branches are just summed up.
The inbalance  of pressure, due to the quasiparticles, is compensated by the
elastic forces in the medium.

\begin{figure}[t]
\begin{center}
\includegraphics[viewport=96 280 500 560, width=0.6\textwidth]{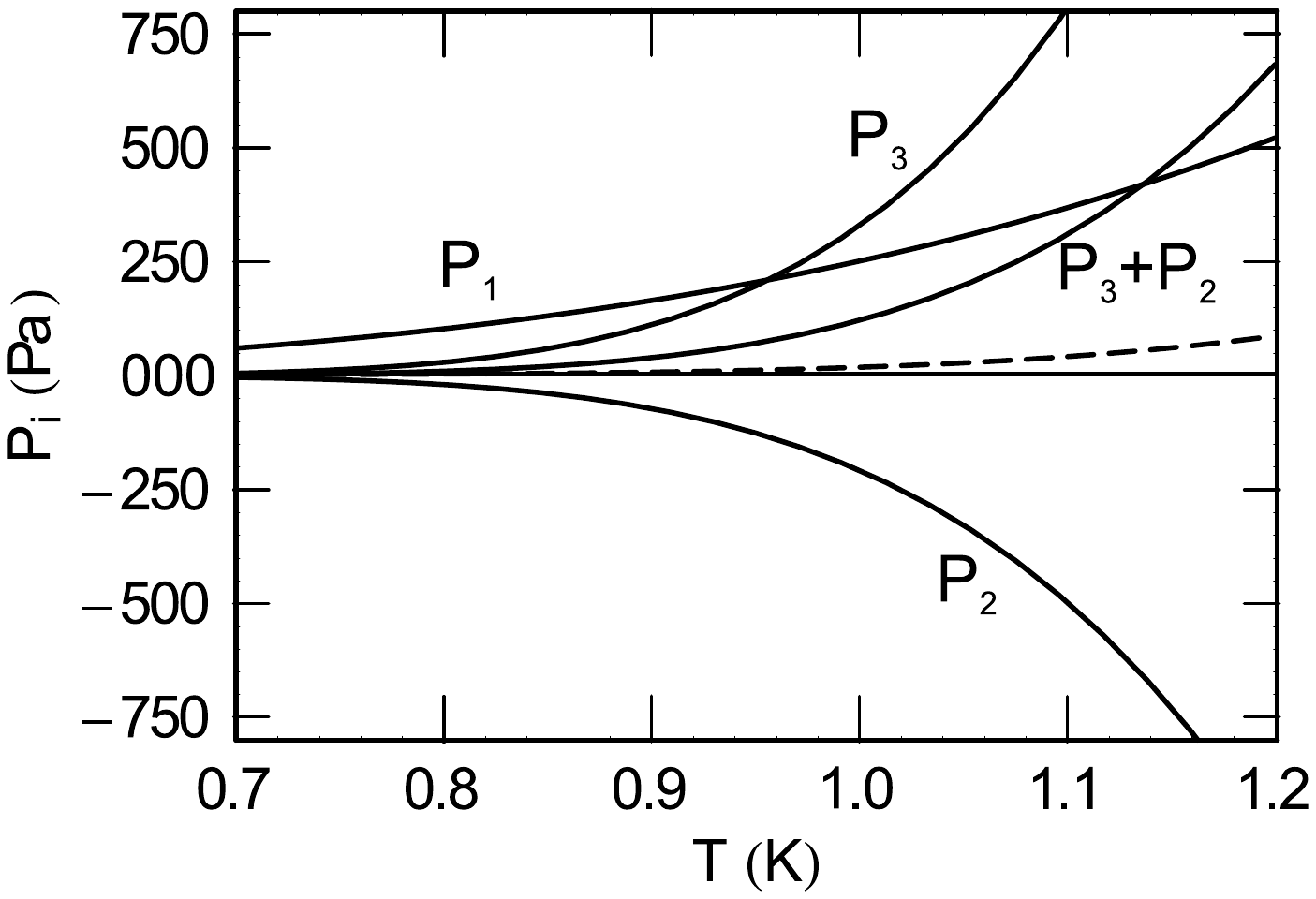}
\parbox{0.9\textwidth}
	{\caption{\label{Fig3}The calculated partial pressures $P_{1,2,3}$ and $P_{3}+P_{2}$ from
Eqs.  (\ref{P1appr})-(\ref{P23appr}). Note that the  absolute value of the
negative pressure due to the $R^-$ rotons is several times greater than the
total roton pressure
$P_{3}+P_{2}$. For comparison the saturated vapour pressure is shown by the
dashed line.}}
\end{center}
\end{figure}

The sound velocity of  the solid is usually much greater than that of
superfluid helium,
  so the pressure of
phonons in the solid is negligibly small (so the trivial extension of the model
including transverse phonons is unnecessary). The partial pressures
of the helium
quasiparticles $P_i$ for $i\!=\!1,2,3$ can be calculated with the usual
approximation of the dispersion relation: linear for phonons and parabolic for
rotons (as in (\ref{phononF}) and (\ref{rotonF})). Then for rotons we have
$p_{2,3}\!=\!p_{rot}\mp\sqrt{2\mu(\varepsilon\!-\!k_B\Delta_{rot})}$, and
considering the second term small, we obtain
\begin{eqnarray}
          \label{P1appr}
      P_{ph}&=&P_{1}=
      \frac{\pi^2}{90}\frac{(k_{B}T)^4}{\hbar^{3}s^{3}},\\
          \label{P23appr}
      P_{2,3}&=&\frac{p_{rot}^{3}}{6\pi^{2}\hbar^{3}}
      \left\{
      \mp k_{B}T
      +3\frac{\sqrt{\pi\mu/2}}{p_{rot}}(k_{B}T)^{3/2}
      \right\}
      e^{-\Delta_{rot}/T}.
\end{eqnarray}
The upper sign (minus) here corresponds to the $R^{-}$ rotons ($i\!=\!2$), and the lower sign (plus) to the $R^{+}$ rotons ($i\!=\!3$). The sum of the pressures in Eqs. (\ref{P1appr}) and (\ref{P23appr}) give the
pressure deduced from thermodynamics in Eq. (\ref{pressure}).
The contributions of
phonons,  $R^{-}$ and $R^{+}$ rotons to pressure are shown in Fig.\ref{Fig3}. For
temperatures $T\!\sim\!1$ K the second term in (\ref{P23appr}) is about four
times smaller than the first one. So the separate contributions of $R^{+}$ and
$R^{-}$ rotons, that differ in sign, are much larger than their sum
$|P_{2,3}|\!>\!P_{rot}\!=\!P_{2}\!+\!P_{3}$.

The measured  saturated vapour
pressure may be  expressed to a good approximation as, \cite{Wyatt1984},
\begin{equation}
     \label{SVP}
     P_{SVP}=
     \left(\frac{m}{2\pi}\right)^{3/2}
     \frac{(k_{B}T)^{5/2}}{\hbar^{3}}
     e^{-E_{B}/T},
\end{equation}
where $m$ is the helium atom mass, $E_{B}\!=7.16$ K the binding energy of
an atom in the liquid (see Fig.\ref{Fig3}). The Eq. (\ref{SVP}) describes well the experimental data for temperatures from $5\cdot10^{-4}$ K to $\sim 1$ K.

Comparing Eqs. (\ref{P23appr}) and
(\ref{SVP}), we get
\begin{equation}
     \label{Psvp-R}
     \frac{P_{SVP}}{P_{rot}}=\sqrt{\frac{m}{\mu}}\frac{k_{B}T}{2p_{rot}^{2}/m}
     e^{(\Delta_{rot}-E_{B})/T}.
\end{equation}
This ratio is at most $0.1$ at $T\sim1 $ K and decreases exponentially at lower
temperatures, where the phonon contribution to pressure becomes dominant.
So in the full temperature range the saturated vapour pressure is at least an
order of magnitude less than the  pressure of quasiparticles (see
Fig.\ref{Fig3}).  This means that the liquid is in tension at the saturated
vapour pressure.

\subsection{Contributions of $R^-$ and $R^+$ rotons in thermodynamics}
The separate contributions of  the $R^+$ rotons and the $R^-$ rotons
cannot be obtained from thermodynamics.
The contributions to the free energy in Eq. (\ref{F}) for $R^+$
rotons and $R^-$ rotons,
are approximately equal. They both give an approximately equal and
positive contribution to the pressure.
This is in stark contrast to the results
of the kinematic analysis, which shows that the contributions to the pressure
of the $R^-$ rotons and $R^+$ rotons are negative
and positve respectively, and the modulus of the pressure due
to the $R^+$ rotons is larger than that from the $R^-$ rotons.
However, the separate contribution
to the pressure from the phonons,
is the same by thermodynamics and kinematics.

For the same reason, the separate contributions of  the $R^+$ rotons
and the $R^-$ rotons
to the fountain pressure, cannot be calculated from their separate
contributions to the entropy, which are both positive.
However the contribution of the phonons and all the rotons can be separated.
The increments of
pressures $\delta P_{1}$ and $\delta P_{rot}$ that are caused by the increment
of temperature $\delta T$
\begin{equation}
      \label{dP}
      \delta P_{1}\!=\!S_{ph}\delta T;\quad \delta P_{rot}\!=\!S_{rot}\delta T,
\end{equation}
where $S_{ph}$ and $S_{rot}$ are the entropies per unit of volume of a phonon
gas
$S_{ph}$ and roton gas $S_{rot}$ respectively, obtained from (\ref{phononF})
and (\ref{rotonF}) (see \cite{Khalatnikov}). Then the full increment of
pressure $\delta P$ that corresponds to $\delta T$ is $\delta P\!=\!S\delta T$,
where $S\!=\!S_{ph}\!+\!S_{rot}$ is the entropy of superfluid helium.

The partial fountain pressures $P_{2,3}$ can only be obtained
from $S_{2,3}$ if a term is added and subtracted from $\delta P_{rot}/\delta
T$.
This term is the derivative of the leading term of Eq. (\ref{P23appr}). It is
added to obtain $P_{3}$ and subtracted to obtain $P_{2}$, thus giving
$P_{3}\!>\!0$ and $P_{2}\!<\!0$.

The above reasoning might be an indication that in the general case,
thermodynamics alone cannot separate the contributions of different
branches of a dispersion curve to the thermodynamic quantities,
and in particular for the case of $R^-$ and $R^+$ rotons.

\section{Observation of negative momentum transfer}

\begin{figure}[t]
\begin{center}
\includegraphics[viewport=125 310 470 540, width=0.7\textwidth]{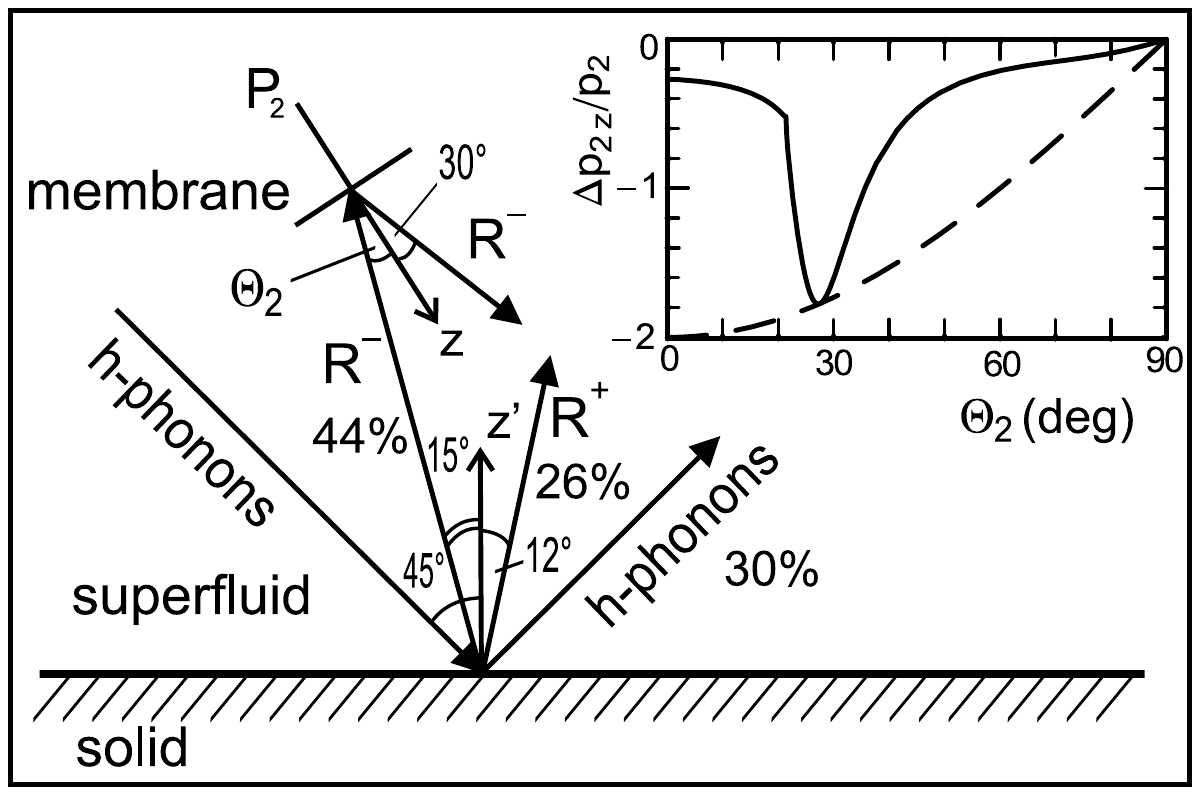}
\parbox{0.9\textwidth}
	{\caption{\label{Fig4}The proposed experimental arrangement for measuring the negative momentum transfer from an incident pulse $R^-$ rotons to an interface (membrane). High energy phonons (h-phonons) are reflected with a mode change into $R^-$ rotons at the first interface, and these $R^-$ rotons are then specularly reflected at the second interface. The angles are chosen to maximise the pressure signal $P_2$, see text. $P_2$ is predicted to be towards the incident beam of $R^-$ rotons. The inset graph shows the ratio of momentum transfer $\Delta p_{2z}$, to the interface per incident $R^-$ roton, to its absolute value of momentum $p_{2}$, as a function of the incidence angle $\Theta_{2}$, (solid line); if the $R^-$ roton were fully
specularly reflected, this function would be $(-2\cos\Theta_{2})$, (dashed line).}}
\end{center}
\end{figure}

We now suggest a way to observe the negative momentum transfer from $R^-$
rotons to an interface, by measuring the pressure caused by a pulse of incident
$R^-$ rotons. The beam of $R^-$ rotons could  be  created by having a pulse of
high-energy helium phonons ($h$-phonons, created in experiments
\cite{exp1,exp2,Hph2,Hph3}), incident on the interface with another solid (see
Fig.\ref{Fig4}). It has been predicted, see \cite{Nasha2008}, that the mode changing
reflection can efficiently produce $R^-$ rotons. The beam of $R^-$ rotons then
should be incident on, for example, a membrane at the optimum angle to
specularly reflect the $R^-$ rotons, see Fig.\ref{Fig4}, as this will give the largest
momentum transfer. The force on the membrane should be outwards.

The created $R^-$ rotons have the same energy as the $h$-phonons, about
$\varepsilon_{h}/k_{B}\!\approx\!10$ K. At $T\sim$ 50 mK both the
$h$-phonons and
the  $R^-$ rotons propagate ballistically (see for example \cite{R-reg} and
\cite{Hphonons}). A fraction of the energy in the h-phonon beam is transferred
to the $R^-$ rotons with probability $R_{12}$. This is predicted (see
\cite{Nasha2008}), to reach $1/2$ at normal incidence (for $\varepsilon_{h}$),
and decrease slowly with increasing angle, so that at $\pi/4$ it is still about
$0.44$. The concentration ratio of the rotons in the reflected beam $n_{rot}$
to the phonons in the incident beam $n_{ph}$ is given by
\begin{equation}
     \label{n-n}
     n_{rot}|u_{rot}|\cos\Theta_{rot}=R_{12}n_{ph}|u_{ph}|\cos\Theta_{ph},
\end{equation}
where $\Theta_{ph}$ and $\Theta_{rot}$ are the incidence and reflection angles,
$u_{ph}$ and $u_{rot}$ group velocities of the quasiparticles at
$\varepsilon_{h}$. The angles are related through a modification of Snell's law
$\sin\Theta_{rot}/s_{2}\!=\!\sin\Theta_{ph}/s_{1}$, where $s_{1,2}$ are the
phase velocities of phonons and $R^-$ roton at energy $\varepsilon_{h}$.

The pressure $P_{m}$, which is created by the $R^-$ roton pulse acting on the
interface, when incident at angle $\Theta_{2}$ to the normal $z$, is determined
by $\Delta p_i$ from Eq. (\ref{momFull}), multiplied by the number of $R^-$
rotons incident per unit time per unit area:
\begin{equation}
     \label{Pexp}
     P_{m}=n_{rot}|u_{rot}|\cos\Theta_{2}\Delta p_{2}.
\end{equation}
In contrast to the isotropic pressure $P_{i}$ from Eq. (\ref{Pi3D}), $\Delta
p_i$ depends strongly on the explicit functions $R_{ij}$, as the reflection of
other types of quasiparticles reduces the negative force. So the transfer of
negative momentum reaches its maximum value when only $R^-$ rotons are
reflected, with no phonons or $R^+$ rotons. Then $\Delta
p_{2}\!\approx\!-2p_{2z}$. It appears
(see \cite{Nasha2008}), that there is a range of incidence angles where
$R_{22}$ is nearly unity, and $R_{21,23}$ nearly zero, see inset of Fig.\ref{Fig4}.

Taking into account the expression for the energy density flux in the incident
beam of h-phonons
\begin{equation}
     \label{Qh}
     Q_{h}=\varepsilon_{h} n_{ph}|u_{ph}|,
\end{equation}
we can use Eqs. (\ref{n-n}) and (\ref{Pexp}) to express the pressure on the
membrane in the form
\begin{equation}
     \label{PexpQ}
     P_{m}=\frac{Q_{h}}{s_{2}}
     \cdot R_{12}\frac{\cos\Theta_{ph}}{\cos\Theta_{rot}}
     \cdot \frac{\Delta p_{2}}{p_{2z}}\cos^{2}\Theta_{2}.
\end{equation}

It can be measured directly in the experiment. The scheme for the experiment is shown in Fig.\ref{Fig4}. The angular dependence of the
transferred momentum to the interface per incident $R^-$ roton $\Delta
p_{2}/p_{2z}$ for energy $\varepsilon_{h}$, with the probabilities taken from
\cite{Nasha2008}, is shown inset in Fig.\ref{Fig4}. It is negative for all angles, and
tends to $(-2)$ at the optimal incidence angle $\Theta_{2}\!=\!\Theta_{2}^{0}$.
For energy $\varepsilon_{h}=10$ K, it is $\Theta_{2}^{0}\!\approx\!27^{\circ}$.
For $\Theta_{ph}\!\approx\!45^{\circ}$ we obtain from (\ref{PexpQ}) that
$|P_{m}|\!\approx\!Q_{h}/2s_{2}$. So due to the negative pressure of the $R^{-}$ rotons the membrane should bend in the direction towards the incident beam, and a sharp peak of the curvature exists around angles $\Theta_{2}^{0}$ (see Fig.\ref{Fig4}).

\section{Conclusion}

We have derived the pressure of the gas of superfluid helium quasiparticles on
an interface immersed in helium. The processes of quasiparticles mode change on
the interface are considered self-consistently in a general way. The
contributions to the pressure, due to phonons, $R^-$ rotons, and $R^+$ rotons,
in thermodynamic equilibrium (\ref{Pi3D}) are found. The contribution of the
$R^-$ rotons is shown to be  negative (\ref{P23appr}). It partially compensates
for the positive contribution of the $R^+$ rotons, so that the resulting roton
pressure in equilibrium is always positive (see Fig.\ref{Fig3}).

The partial pressures of quasiparticles of different types (\ref{Pi3D}) can be
expressed in the form of the pressure of a classical gas (\ref{Pclass}),
despite the fact that quasiparticles interact with the interface in a much more
complex manner. This result should hold true for any two adjacent continuous
media, as the explicit forms for the probabilities, of relevant processes, do
not appear in the expression for the pressure. One of the consequences is that
the net  force, which the quasiparticles of both media exert on the interface,
is directed towards the medium with the greater sound velocity of the two. This
inbalance  of pressure, due to the quasiparticles, is compensated by elastic
forces which leads to the liquid being under tension at the saturated
vapour pressure.

It is shown that the  pressure of quasiparticles is that which underlies the
fountain effect in helium (\ref{dP}). So the fountain effect is due to the
osmotic pressure of the quasiparticles that are ``in solution" in the
superfluid. The negativeness of the pressure due to $R^-$ rotons but
with the obvious
positiveness of their
contribution to the entropy of superfluid helium, is explained. We show that
the equation $\delta P=S\,\delta T$ cannot be applied to the $R^{-}$ and
$R^{+}$ parts of the dispersion curve separately, although it appears that it
can be applied separately to the phonons and all the rotons.

An experimental setup is suggested (Fig.\ref{Fig4}) for detecting the negative momentum
transferred to a membrane by a $R^-$ roton beam, which is created by mode
change reflection of $h$-phonons at an interface with a solid. We hope this
paper stimulates experiments to test these predictions.

\begin{acknowledgements}
We are grateful to EPSRC of the UK (grant EP/F 019157/1)
for support of this work.
\end{acknowledgements}

\end{document}